\documentclass[a4paper]{article}
\pdfoutput=1

\usepackage{icrc2013}
\usepackage[english]{babel}

\title{FACT: Towards Robotic Operation of an Imaging Air Cherenkov Telescope}

\shorttitle{FACT Robotic Operation}

\newcommand{\ethz}{$^1$}
\newcommand{\tudo}{$^2$}
\newcommand{\epfl}{$^4$}
\newcommand{\unige}{$^3$}
\newcommand{\uniw}{$^5$}

\newcommand{\uniz}{$^{a}$}
\newcommand{\unizh}{$^{^a}$}
\newcommand{\kynu}{$^{b}$}
\newcommand{\kynuh}{$^{^b}$}
\newcommand{\mpim}{$^{c}$}
\newcommand{\mpimh}{$^{^c}$}
\newcommand{\tum}{$^{d}$}
\newcommand{\tumh}{$^{^d}$}

\authors{
A.~Biland\ethz,
H.~Anderhub\ethz,
M.~Backes\tudo,
V.~Boccone\unige,
I.~Braun\ethz,
T.~Bretz\ethz,
J.~Bu\ss\tudo,
F.~Cadoux\unige,
V.~Commichau\ethz,
L.~Djambazov\ethz,
D.~Dorner\uniw,
S.~Einecke\tudo,
D.~Eisenacher\uniw,
A.~Gendotti\ethz,
O.~Grimm\ethz,
H.~von Gunten\ethz,
C.~Haller\ethz,
D.~Hildebrand\ethz,
U.~Horisberger\ethz,
B.~Huber\ethz\unizh,
\mbox{K.-S.}~Kim\ethz\kynuh,
M.~L.~Knoetig\ethz,
\mbox{J.-H.}~K\"ohne\tudo,
T.~Kr\"ahenb\"uhl\ethz,
B.~Krumm\tudo,
M.~Lee\ethz\kynuh,
E.~Lorenz\ethz\mpimh,
W.~Lustermann\ethz,
E.~Lyard\unige,
K.~Mannheim\uniw,
M.~Meharga\unige,
K.~Meier\uniw,
T.~Montaruli\unige,
D.~Neise\tudo,
F.~Nessi-Tedaldi\ethz,
\mbox{A.-K.}~Overkemping\tudo,
A.~Paravac\uniw,
F.~Pauss\ethz,
D.~Renker\ethz\tumh,
W.~Rhode\tudo,
M.~Ribordy\epfl,
U.~R\"oser\ethz,
\mbox{J.-P.}~Stucki\ethz,
J.~Schneider\ethz,
T.~Steinbring\uniw,
F.~Temme\tudo,
J.~Thaele\tudo,
S.~Tobler\ethz,
G.~Viertel\ethz,
P.~Vogler\ethz,
R.~Walter\unige,
K.~Warda\tudo,
Q.~Weitzel\ethz,
M.~Z\"anglein\uniw $\;\;$
(FACT Collaboration)
}
\afiliations{
\ethz ETH Zurich, Switzerland $\;\;-\;\;$
Institute for Particle Physics, Schafmattstr.~20, 8093 Zurich\\
\tudo Technische Universit\"at Dortmund, Germany $\;\;-\;\;$
Experimental Physics 5, Otto-Hahn-Str.~4, 44221 Dortmund\\
\unige University of Geneva, Switzerland $\;\;-\;\;$
ISDC, Chemin d'Ecogia~16, 1290 Versoix $\;\;-\;\;$
DPNC, Quai Ernest-Ansermet 24, 1211 Geneva\\
\epfl EPF Lausanne, Switzerland  $\;\;-\;\;$
Laboratory for High Energy Physics, 1015 Lausanne\\
\uniw Universit\"at W\"urzburg, Germany $\;\;-\;\;$ 
Institute for Theoretical Physics and Astrophysics, Emil-Fischer-Str.~31, 97074 W\"urzburg\\
\scriptsize{
\uniz {Also at: University of Zurich, Physik-Institut, 8057 Zurich, Switzerland}\\
\kynu {Also at: Kyungpook National University, Center for High Energy Physics, 702-701 Daegu, Korea}\\
\mpim {Also at: Max-Planck-Institut f\"ur Physik, D-80805 Munich, Germany}\\
\tum  {Also at: Technische Universit\"at M\"unchen, D-85748 Garching, Germany}\\
}
}

\email{biland@phys.ethz.ch}

\abstract{
The First G-APD Cherenkov Telescope (FACT) became operational at La Palma
in October 2011. Since summer 2012, due to very smooth and stable operation,
it is the first telescope of its kind that is routinely operated from
remote, without the need for a data-taking crew on site. In addition, many
standard tasks of operation are executed automatically without the need for
manual interaction.
Based on the experience gained so far, some alterations to improve the
safety of the system are under development to allow robotic operation in the
future.
We present the setup and precautions used to implement remote
operations and the experience gained so far, as well as the work towards
robotic operation.
}
\keywords{FACT, IACT, remote operation, robotic operation}

\begin{document}
\maketitle

%Begin a section.
\section{Introduction}

So far, all Imaging Air Cherenkov Telescopes use Photomultiplier
tubes to measure the dim flashes of Cherenkov light emitted by air-showers
induced by high energy cosmic-ray particles or gamma-rays.
The First G-APD Cherenkov Telescope (FACT) investigates the feasibility
of using solid state photosensors (Avalanche Photo Diodes operated in 
Geiger-mode: G-APD aka SiPM) for future cameras. A complete camera
consisting of 1440 pixels was designed, constructed and installed
in the refurbished HEGRA CT3 telescope, having a mirror area of
$\approx 9.5$ m$^2$,  at the canary Island La Palma
next to the two huge MAGIC telescopes \cite{JINST,Bretz}.

The G-APDs
are type Hamamatsu MPPC S10362-33-50C, equipped with solid light concentrators.
The electronics is based on the DRS4  chip and read via standard
Ethernet. The physics trigger uses analogue sums of nine neighbouring
pixels forming a trigger-patch, and the complete camera is read out
in case one patch has a signal above a programmable threshold.
The complete trigger and data-acquisition electronics is integrated in
the camera body.

The power supplies for the electronics as well as bias voltage for the
G-APDs are located in a container next to the telescope, housing also
the control cabinet for the drive systems, the computers for
data-acquisition, control, onsite data-analysis and communication as
well as working space for an onsite shift crew.

There are several individual processes to operate the telescope, mainly
drive control, slow control, trigger control, bias control and
daq control, all steered by a central control. All the communications
between these systems uses the DIM environment\cite{DIM}, allowing
also easy integration of different user interfaces.
Currently there exists a command-line interface, a full graphical
interface allowing also full debugging of the system, and a
reduced graphical interface for standard operations
that can be run from smartphones.

Power is delivered from the MAGIC system, including a large UPS and
a diesel generator in case of extended power outages. Internet
is based on the system delivered by the ORM (Observatorio
Roque de los Muchachos).

\section{Early Operation and Identified Problems}

The FACT camera was mounted in October 2011, and few hours after
cabling the first airshowers were successfully recorded 
(under fullmoon conditions). While it was planned
to bring the camera back to the laboratory for some upgrade
work after few months of experience gained, the system is
working so reliable since the first minute that this upgrade
was canceled and FACT is taking data since then almost
every night.

Very few problems have been encountered so far, mainly when
powering on the camera. It can happen that some components do
not correctly boot, but all these problems can be solved by
a reset command or a power cycling of the camera electronics.

During operation, it can happen that some communications get lost.
All these can be identified within very short time, and a simple
reconnect command was always sufficient to restore operation.

Few incidences needed a manual interaction when a FACT
crew was still on site:\\
%\begin{itemize}
% \item
-- Once the drive system behaved erratically and it was necessary
   to park the telescope manually. This was traced to a cable
   in the control cabinet of the drive system being damaged
   and was repaired within two days. \\
-- Recently, the electronics of the drive system started to suffer
   from very high temperatures in the container. We are going to install
   a cooling system.
% \item
-- A fuse inside the camera housing protecting the electronics
   did blow. The cause might have been a DRS4 not
   correctly initializing and therefore consuming far too much
   power. In the mean time, additional monitoring is installed
   to prevent such an incident. Being
   a prototype, some necessary changes in the basic design
   during construction
   resulted in a crowded system not allowing easy maintenance
   of the electronics inside the camera housing. Nevertheless,
   a procedure was developed at site and the damage repaired
   within few days. \\
%\item
-- The electronics controlling the Bias supply lost its
   firmware for unknown reasons and was replaced by a spare board.\\
%\item
-- One of the boards delivering Bias voltages did break and
   was replaced by a spare board. \\
-- After almost two years of operation, the water pump for 
   cooling the camera electronics needed a major maintenance.

%\end{itemize}

Beyond these, the only manual interactions ever necessary were
refilling of cooling water, greasing the hinges of the
camera lid and replacing a broken disk in a RAID array.
Recently, we suffered for few nights from overheating of the
electronics of the drive system and are improving
its cooling.

The stability and reliability of the system
allows to operate FACT since spring 2012 from remote, without
the need of a shift-crew permanently at La Palma. Currently, we
are working on minor auxiliary system improvements to allow robotic
operation in near future.
Of course, the system cannot be left fully unattended,
and we have an agreement with the MAGIC collaboration to get
some help from the onsite crew in case of emergencies.
But the need for help shall be limited to exceptional cases.

\section{Personnel safety}

By far the most important aspect of operation of a system without
onsite crew is to care for safety of people who could be 
near to the telescope.
Therefore, the
telescope is fully fenced without a gate. The only possibility to
enter is through the container, having one door outside and one
inside of the gate. The container is locked, with the MAGIC
crew having access to the key in case emergency access is needed.
If work has to be performed at the telescope structure, a large 
switch disables the power to the drive system.

A infrared CCD can monitor the area. If it is too dark,
a infrared illumination can be activated, but this has to be
limited to emergencies since it does affect datataking.

\section{Remote operation}

Not only for remote, but also
for onsite operation it is a good strategy
that major hardware components shall protect themself against
malfunctioning. Section 6 will explain in detail how
this is implemented for the major subsystems.

This leaves as key difference between onsite and remote operation
that it is not possible to press buttons on e.g. front panels
of power supplies or switching power off and on. Fortunately the
AGILENT \cite{agil} power supplies used can be fully controlled via Ethernet,
needing no modifications. Minor systems like the lid control
were upgraded with simple Ethernet connectivity based on
Arduino \cite{ard} boards. In addition, we want to be able to power cycle
each component via its 220V connection. Therefore, we installed
Ethernet controllable power switches (GUDE 8012 \cite{gude}), and did
configure all computers to boot at power up. For security reasons,
all Ethernet connections are limited to a internal network, with
the only connection from outside via two redundant gateway computers.

This leaves a major vulnerability: if both gateway computers are
stuck, there is no possibity to access the power switch to reboot
them. Therefore, the computers are connected to a power switch
GUDE 8090 \cite{gude} that is not only accessible via the internal network,
but also via sending a SMS message from a cellular phone via GSM.
This allows to power cycle all
components necessary for Internet communication from remote, independent
of the Internet.

Since more than one year we operate FACT from remote without a
major hardware problem. During this time, we upgraded the software
to identify and cure every incident we encountered so far. In
addition, all operations except the safety relevant parts like
unlocking the drive system are now executed via scripts,
configuration files and database entries. This reduces the duty of
the remote shifter more and more to a simple monitoring task.

\section{Robotic Operation}

The step from remote to robotic operation is mainly to ensure
that the system is able to operate on its own and can
handle all the known or foreseeable malfunctions. Save operation
shall have priority over extended datataking, so in case of
an unexpected problem the system should put itself into a save
state (i.e. at least parking the telescope and switching off the 
bias voltage of the G-APDs) and inform the collaboration by
e.g. sending an e-mail, SMS or automatic phone calls.
The most dangerous situation is the telescope not in
the park position when the sun rises, since we could
concentrate more than 10 kW solar power onto an area of
 few cm$^2$ on the camera.

We are currently testing the operation software and ensuring
the system to behave correctly under all known conditions. Some
minor auxiliary systems still need to be adapted. When this is
done, we will send a crew to La Palma to manually trigger all
expected malfunctions and check if the system reacts correctly.

\section{Major Subsystems}

In this section we describe how the major subsystems are implemented
and protected. For simplicity,
the descriptions concentrate mainly on the topology and
do not necessarily represent the detailed implementation. Due to
usage of off the shelf components and a historically grown system, some
tasks are more complicated than necessary and could be simplified
in a future implementation. We plan to have all systems ready
for robotic operations by late 2013.

We describe here only the safety related tasks and leave out the
normal dataflow and the datataking procedures working on top of it.

\subsection{Electronics Power}

\begin{figure}[ht]
  \centering
  \includegraphics[width=0.475\textwidth]{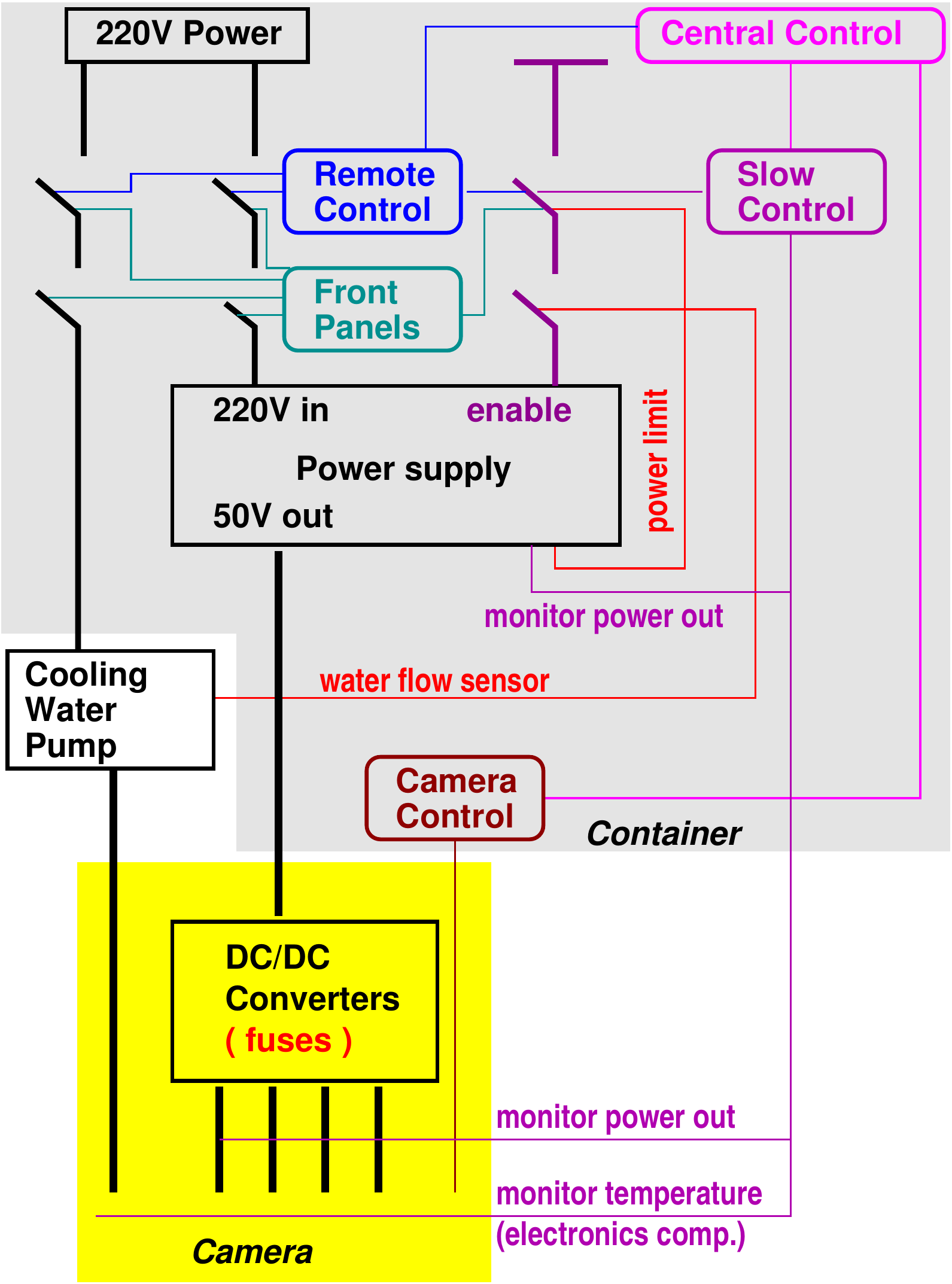}
  \caption{Schematics of the electronics power control. See 6.1
   for explanations.}
  \label{power}
 \end{figure}

A schematic of the camera power and safety setup is shown in
figure \ref{power}.

The key component is the central power supply delivering 50V
to the electronics integrated in the camera body.
While the operation power consumption is less than 500W, the high
density of the electronics still asks for continuous cooling.
We use a water cooling, and whenever the water flow stops,
the power supply is automatically disabled. Without making hardware
shortcuts it is therefore not possible to deliver power to the
camera without operational water cooling.

In addition, the power supply is configured with an internal
power limit. Unfortunately, due to high spikes in power consumption
during booting of the system, this limit has to be set rather high.
Therefore, we are monitoring the delivered power, and the slow
control system shall disable the power supply when the delivery is
above a predefined limit
for several seconds. The slow control system monitors the
power output of the DC/DC converters inside the camera housing
(as a last resort, these are equipped with fuses) delivering the
power to the individual electronics boards. It also measures the
temperature inside the electronics compartment of the camera.
If these are above predefined limits, the power supply will be
disabled.
Such disable states can be reset via the front
panel or from remote control via Ethernet
when the cause of the problem has been resolved.

The 220V power for the cooling pump can be switched from a front
panel or the remote control concurrently, e.g. it is possible
to switch the power off locally and switch it on again from
remote.

For the power supply, it is also possible to switch the 220V
concurrently from local or remote, but there is the additional
possibility to switch it off locally without the possibility
for remote activation. This allows to ensure the power is off
in case of a crew working onsite on the camera.

What is summarized in the figure as camera control includes trigger control
and data acquisition. All control task communicate with the central
control.

\subsection{Bias Power}

\begin{figure}[ht]
  \centering
  \includegraphics[width=0.475\textwidth]{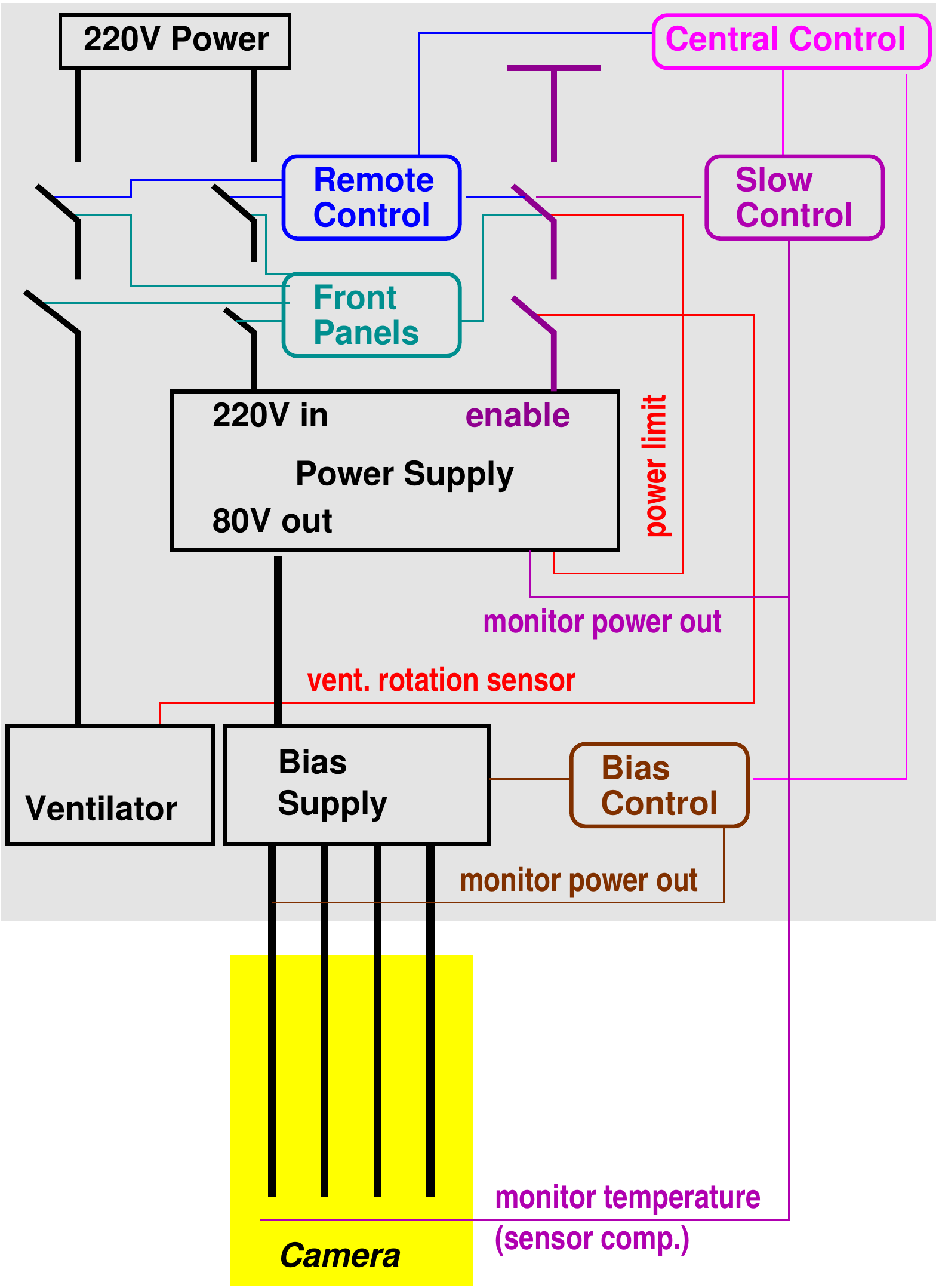}
  \caption{Schematics of the Bias power control. See 6.2
   for explanations.}
  \label{fig2}
 \end{figure}

In addition to the power for the electronics, there is also the
need to deliver the bias power to the G-APDs (fig.\ref{fig2}).
Ensuring the correct
voltage setting is the task of the Bias control. It does also
monitor the total power delivered and shall set the voltage to
zero in case of too high power consumption.

The control of the  power supply delivering 80V is
configured the same way as described in the previous section. The
only difference is there is no water cooling but a ventilator
that is necessary to cool the bias supply. To ensure the blades
are turning, we have glued small permanent magnets on the blades
and read hall sensors positioned on top of them. In case some
of the blades do not turn, the power supply is disabled and
it is not possible to overrule that from remote.
The ventilators can also be switched on and off concurrently
from local or remote, and a dedicated local power switch
exists but shall only but used in case somebody is
working directly at the bias supply onsite.

The slow control system is monitoring the total power delivered by
the power supply and the temperature in the sensor compartment of
the camera. It shall disable the power if the values are above
a predefined limit.

\subsection{Drive System}

The drive system can be locally operated with a joystick and the
power can be switched on and off from a large manual switch,
The joystick also includes an emergency button to break the
power to the motors of the drive system.

\begin{figure}[ht]
  \centering
  \includegraphics[width=0.475\textwidth]{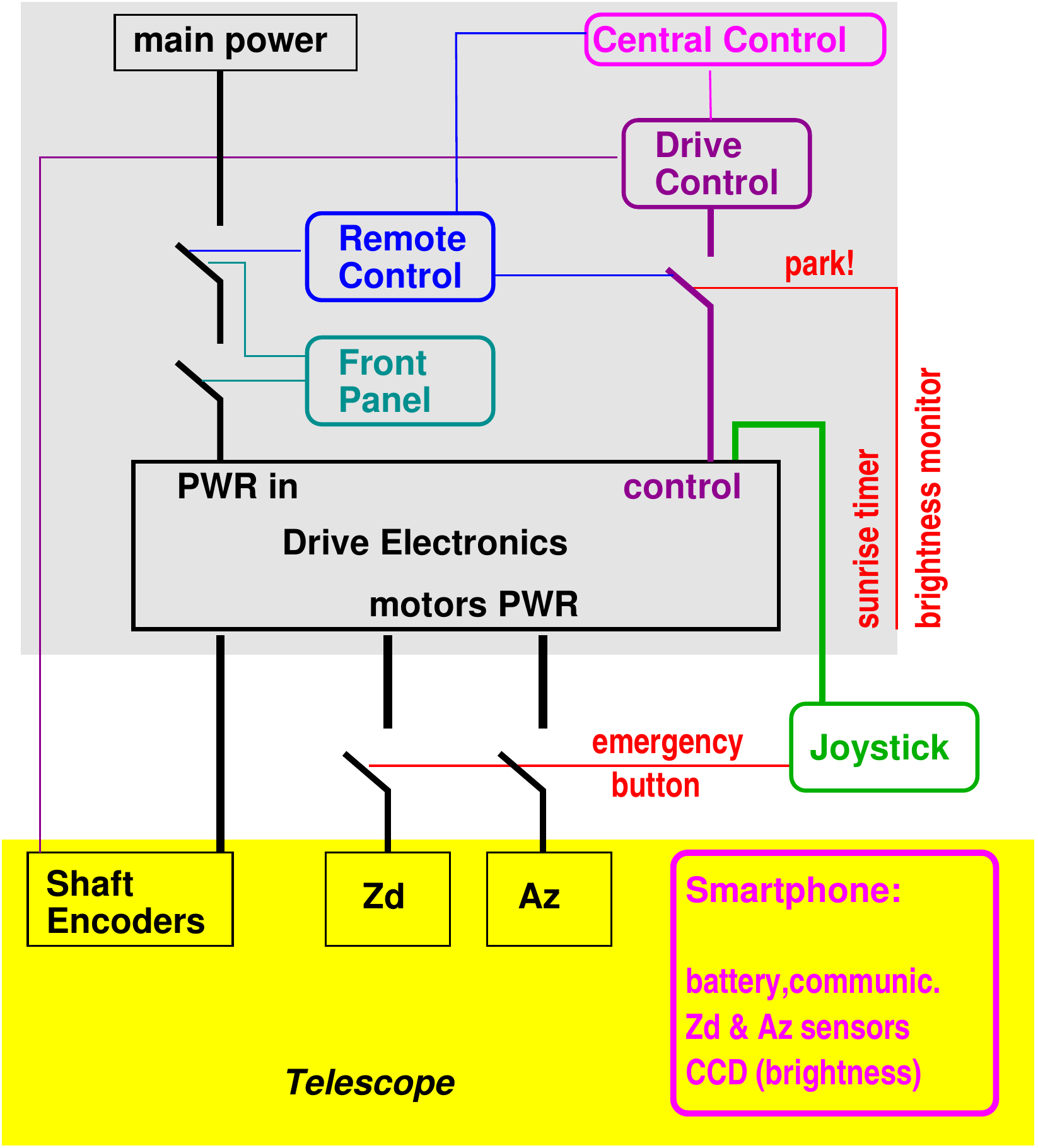}
  \caption{Schematics of the drive system control. See 6.3
   for explanations.}
  \label{fig3}
 \end{figure}

Normal operation (Fig \ref{fig3}), nevertheless, are done via the drive control
sending commands to the drive electronics and reading the shaft
encoders to know the orientation of the telescope.
While the drive control is responsible to park the
telescope before the sun rises, there is an additional brightness
monitor and and a sunrise-timer that will send a 'park' command
and disable the drive control before there is danger to the
telescope. The drive control cannot overrule these conditions
itself. A dedicated command from the remote control is needed
to enable the drive control again.

In addition to the main manual power switch that cannot be
overruled from remote, there is the default power switch
that can be used from local or remote concurrently.

Ensuring the parking of the telescope is the most crucial part
of robotic operation. But there is a major problem that it might
not be possible to check the pointing of the telescope from
remote  for several reasons like: \\
%\begin{itemize}
% \item
-- Due to a major power problem it is not possible to access
   the shaft encoders or the computers reading the shaft encoders.\\
% \item
-- A local Ethernet problem or a Internet problem to La Palma
   prevents accessing the computers reading the shaft encoders.\\
% \item
-- All computers are down and connection cannot be established
   in due time.                                             
%\end{itemize}

In case of a power or network failure, most probably also the
installed webcam cannot be used to check the orientation of
the telescope.
Therefore, we are currently investigating a fully independent
system, that needs to have its own power, independent communication
and sensors to approximately read the zenith and azimuth position
of the telescope. Most probably this can be reached by implementing
a standard smartphone in the telescope structure: the batteries
ensure independent power for several hours, it knows the actual time,
B-field sensor
and inclinometer measure the orientation, and by definition it
can use  GSM for communication.
Additionally, the integrated CCD camera might be used as a
brightness sensor or used to take pictures of the environment.
This smartphone can then be queried from remote,
or independently send alerts if the telescope is not parked in
due time. It is then possible to park the telescope manually
before it can be damaged.

\subsection{Documentation}

When operating a robotic telescope, it can not be expected that
information is flowing from one shift crew to the other. In
addition, if problems are rare theire solution tend to get
forgotten. Therefore, it is even more crucial than for normal
telescopes to have a well and up to date documentation available
on site as well as accessible from remote places.
We must have online versions on the computers on site and mirrored
to several machines at the institutes and laptops of the users
as well as a printed version in the container.

\section{Conclusion}

FACT is working very reliable since being switched on the first
time in October 2011. The majority of the few hardware
related problems encountered so far did not need any local
interaction but can be solved by software, allowing to operate
the system from remote. Based on the experience of more than
one year remote operation, handling of problems was more and
more automatized, and the operation of the telescope reduces
to monitoring and identifying bugs in the control software.
This will eventually lead to a fully robotic operation, only
needing help from onsite personnel in case of severe emergencies.

%\vspace*{0.5cm}
\vfill

\footnotesize{
\paragraph{Acknowledgment} {
The important contributions from ETH Zurich grants ETH-10.08-2 and
ETH-27.12-1 as well as the funding by the German BMBF (Verbundforschung
Astro- und Astroteilchenphysik) are gratefully acknowledged. 
% We are thankful for the very valuable contributions from E. Lorenz,
% D. Renker and G. Viertel during the early phase of the project
We thank the
Instituto de Astrofisica de Canarias allowing us to operate the telescope
at the Observatorio Roque de los Muchachos in La Palma, and the
Max-Planck-Institut f\"ur Physik for providing us with the mount of the
former HEGRA CT\,3 telescope,
and the MAGIC collaboration for their support.
We also thank the group of Marinella Tose 
from the College of Engineering and Technology at Western Mindanao
State University, Philippines, for providing us with the scheduling
web-interface.
 }}

\end{document}